%% file: science_template.tex
\documentclass[12pt]{article}

\usepackage{newtxtext,newtxmath}

\usepackage{graphicx}

\usepackage[letterpaper,margin=1in]{geometry}

\newcommand{\snname}{SN~Eos}

\newcommand{\webb}{\textit{JWST}}
\newcommand{\hst}{\textit{HST}}
\newcommand{\clname}{MACS\,1931.8-2635}

\newcommand{\LCDM}{$\Lambda$CDM}
\newcommand{\lya}{Ly$\alpha$}
\newcommand{\lyalong}{Lyman-$\alpha$}
\newcommand{\halpha}{H$\alpha$}
\newcommand{\hbeta}{H$\beta$}
\newcommand{\hgamma}{H$\gamma$}
\newcommand{\hdelta}{H$\delta$}
\newcommand{\spectralphase}{$\sim 98$~days}
\def\arcsec{\ensuremath{^{\prime\prime}}}

\renewenvironment{abstract}
	{\quotation}
	{\endquotation}

\date{}

\renewcommand\refname{References and Notes}

\makeatletter
\renewcommand{\fnum@figure}{\textbf{Figure \thefigure}}
\renewcommand{\fnum@table}{\textbf{Table \thetable}}
\makeatother

\usepackage{scicite}

\usepackage{url}

\def\scititle{
	A spectroscopically confirmed, strongly lensed, metal-poor Type II supernova at z = 5.13
}
\title{\bfseries \boldmath \scititle}

\author{
	David~A.~Coulter$^{1,2\ast\dagger}$,
	Conor~Larison$^{2\ast\dagger}$,
	Justin~D.~R.~Pierel$^{2}$,
	Seiji~Fujimoto$^{3,4}$,\and
	Vasily~Kokorev$^{5,6}$,
	Joseph~F.~V.~Allingham$^{7}$,
	Takashi~J.~Moriya$^{8,9,10}$,
	Matthew~Siebert$^{2}$,\and
	Yoshihisa~Asada$^{3}$,
	Rachel~Bezanson$^{11}$,
	Maru\v{s}a~Brada\v{c}$^{12,13}$,
	Gabriel~Brammer$^{14}$,\and
	John~Chisholm$^{5,6}$,
	Dan~Coe$^{2}$,
	Pratika~Dayal$^{3,15,16}$,
	Michael~Engesser$^{2}$,\and
	Steven~L.~Finkelstein$^{5,6}$,
	Ori~D.~Fox$^{2}$,
	Lukas~J.~Furtak$^{5,6}$,
	Anton~M.~Koekemoer$^{2}$,\and
	Thomas~Moore$^{2}$,
	Minami~Nakane$^{17,18}$,
	Masami~Ouchi$^{8,17,9,19}$,
	Richard~Pan$^{20}$,\and
	Robert~Quimby$^{21,19}$,
	Armin~Rest$^{2,1}$,
	Johan~Richard$^{22}$,
	Luke~Robbins$^{20}$,\and
	Louis-Gregory~Strolger$^{2}$,
	Fengwu~Sun$^{23}$,
	Tommaso~Treu$^{24}$,
	Hiroto~Yanagisawa$^{17,18}$,\and
	Abdurro'uf$^{25}$,
	Aadya~Agrawal$^{26}$,
	Ricardo~Amor\'{i}n$^{27}$,
	Joseph~P.~Anderson$^{28}$,\and
	Rodrigo~Angulo$^{1}$,
	Hakim~Atek$^{29}$,
	Franz~E.~Bauer$^{30}$,
	Larry~D.~Bradley$^{2}$,
	Volker~Bromm$^{5,6}$,\and
	Mateusz~Bronikowski$^{31}$,
	Christopher~J.~Conselice$^{32}$,
	Christa~DeCoursey$^{33}$,\and
	James~M.~DerKacy$^{2}$,
	Guillaume~Desprez$^{34}$,
	Suhail~Dhawan$^{35}$,
	Jose~M.~Diego$^{36}$,\and
	Eiichi~Egami$^{33}$,
	Andreas~Faisst$^{37}$,
	Brenda~Frye$^{33}$,
	Sebastian~Gomez$^{5}$,\and
	Mauro~Gonz\'alez-Otero$^{27}$,
	Massimo~Griggio$^{2}$,
	Yuichi~Harikane$^{17}$,
	Kohei~Inayoshi$^{38}$,\and
	Saurabh~W.~Jha$^{39}$,
	Yolanda~Jim\'enez-Teja$^{27,40}$,
	Jeyhan~S.~Kartaltepe$^{41}$,
	Patrick~L.~Kelly$^{42}$,\and
	Lindsey~A.~Kwok$^{43}$,
	Zachary~G.~Lane$^{44}$,
	Xiaolong~Li$^{1}$,
	Ivo~Lobbe$^{45}$,
	Paulo~A.~A.~Lopes$^{46}$,\and
	Ray~A.~Lucas$^{2}$,
	Georgios~E.~Magdis$^{14,47}$,
	Nicholas~S.~Martis$^{12}$,
	Jorryt~Matthee$^{48}$,\and
	Ashish~K.~Meena$^{49}$,
	Rohan~P.~Naidu$^{50}$,
	Ga\"{e}l~Noirot$^{2}$,
	Masamune~Oguri$^{51,52}$,\and
	Estefania~Padilla~Gonzalez$^{1}$,
	Massimo~Pascale$^{24}$,
	Tanja~Petrushevska$^{31}$,
	Massimo~Ricotti$^{53}$,\and
	Daniel~Schaerer$^{54}$,
	Stefan~Schuldt$^{55,56,57}$,
	Melissa~Shahbandeh$^{2}$,
	William~Sheu$^{24}$,\and
	Koji~Shukawa$^{1}$,
	Akiyoshi~Tsujita$^{58}$,
	Eros~Vanzella$^{59}$,
	Qinan~Wang$^{50}$,
	John~Weaver$^{50}$,\and
	Robert~Williams$^{2,60}$,
	Rogier~Windhorst$^{61}$,
	Yi~Xu$^{14}$,
	Yossef~Zenati$^{1,62,63}$,
	Adi~Zitrin$^{7}$\and
	\small$^{1}$William H. Miller III Department of Physics and Astronomy, Johns Hopkins University, 3400 North Charles Street,\and
    \small Baltimore, MD 21218, USA.\and
	\small$^{2}$Space Telescope Science Institute, 3700 San Martin Drive, Baltimore, MD 21218, USA.\and
	\small$^{3}$David A. Dunlap Department of Astronomy and Astrophysics, University of Toronto, \and
    \small 50 St. George Street, Toronto, Ontario, M5S 3H4, Canada.\and
	\small$^{4}$Dunlap Institute for Astronomy and Astrophysics, 50 St. George Street, Toronto, Ontario, M5S 3H4, Canada.\and
	\small$^{5}$Department of Astronomy, University of Texas at Austin, 2515 Speedway, Stop C1400, Austin, TX 78712, USA.\and
	\small$^{6}$Cosmic Frontier Center, The University of Texas at Austin, Austin, TX 78712.\and
	\small$^{7}$Department of Physics, Ben-Gurion University of the Negev, P.O. Box 653, Beer-Sheva 8410501, Israel.\and
	\small$^{8}$National Astronomical Observatory of Japan, National Institutes of Natural Sciences, 2-21-1 Osawa, Mitaka, \and 
    \small Tokyo 181-8588, Japan.\and
	\small$^{9}$Graduate Institute for Advanced Studies, SOKENDAI, 2-21-1 Osawa, Mitaka, Tokyo 181-8588, Japan.\and
	\small$^{10}$School of Physics and Astronomy, Monash University, Clayton, VIC 3800, Australia.\and
	\small$^{11}$Department of Physics and Astronomy and PITT PACC, University of Pittsburgh, Pittsburgh, PA 15260, USA.\and
	\small$^{12}$Faculty of Mathematics and Physics, University of Ljubljana, Jadranska ulica 19, SI-1000 Ljubljana, Slovenia.\and
	\small$^{13}$Department of Physics and Astronomy, University of California Davis, 1 Shields Avenue, Davis, CA 95616, USA.\and
	\small$^{14}$Cosmic Dawn Center (DAWN), Niels Bohr Institute, University of Copenhagen, Jagtvej 128, DK2200 \and 
    \small Copenhagen N, Denmark.\and
	\small$^{15}$Department of Physics, 60 St George St, University of Toronto, Toronto, ON M5S 3H8, Canada.\and
	\small$^{16}$Canadian Institute for Theoretical Astrophysics, 60 St George St, University of Toronto, \and 
    \small Toronto, ON M5S 3H8, Canada.\and
	\small$^{17}$Institute for Cosmic Ray Research, The University of Tokyo, 5-1-5 Kashiwanoha, Kashiwa, Chiba 277-8582, Japan.\and
	\small$^{18}$Department of Physics, Graduate School of Science, The University of Tokyo, 7-3-1 Hongo, Bunkyo, \and 
    \small Tokyo 113-0033, Japan.\and
	\small$^{19}$Kavli Institute for the Physics and Mathematics of the Universe (WPI), The University of Tokyo Institutes for \and 
    \small Advanced Study, The University of Tokyo, Kashiwa, Chiba 277-8583, Japan.\and
	\small$^{20}$Department of Physics and Astronomy, Tufts University, 574 Boston Avenue, Suite 304, Medford, MA 02155, USA.\and
	\small$^{21}$Department of Astronomy and Mount Laguna Observatory San Diego State University, San Diego, CA 92182, USA.\and
	\small$^{22}$Univ Lyon, Univ Lyon1, Ens de Lyon, CNRS, Centre de Recherche Astrophysique de Lyon \and 
    \small UMR5574, 69230, Saint-Genis-Laval, France.\and
	\small$^{23}$Center for Astrophysics $|$ Harvard \& Smithsonian, 60 Garden St., Cambridge, MA 02138, USA.\and
	\small$^{24}$Department of Physics \& Astronomy, University of California, Los Angeles, 430 Portola Plaza, \and 
    \small Los Angeles, CA 90095, USA.\and
	\small$^{25}$Department of Astronomy, Indiana University, 727 East Third Street, Bloomington, IN 47405, USA.\and
	\small$^{26}$Department of Astronomy, University of Illinois Urbana-Champaign, 1002 West Green Street, Urbana, IL 61801, USA.\and
	\small$^{27}$Instituto de Astrof\'{i}sica de Andaluc\'{i}a (CSIC), Apartado 3004, 18080 Granada, Spain.\and
	\small$^{28}$European Southern Observatory, Alonso de Córdova 3107, Vitacura, Casilla 19001, Santiago, Chile.\and
	\small$^{29}$Institut d'Astrophysique de Paris, CNRS, Sorbonne Universit\'e, 98bis Boulevard Arago, 75014, Paris, France.\and
	\small$^{30}$Instituto de Alta Investigaci{\'{o}}n, Universidad de Tarapac{\'{a}}, Casilla 7D, Arica, 1010000, Chile.\and
	\small$^{31}$Center for Astrophysics and Cosmology, University of Nova Gorica, Vipavska 11c, 5270 Ajdov\v{s}\v{c}ina, Slovenia.\and
	\small$^{32}$Jodrell Bank Centre for Astrophysics, University of Manchester, Oxford Road, Manchester, UK.\and
	\small$^{33}$Steward Observatory, University of Arizona, 933 N. Cherry Avenue, Tucson, AZ 85721, USA.\and
	\small$^{34}$Kapteyn Astronomical Institute, University of Groningen, P.O. Box 800, 9700AV Groningen, The Netherlands.\and
	\small$^{35}$School of Physics \& Astronomy and Institute of Gravitational Wave Astronomy, University of Birmingham, UK.\and
	\small$^{36}$Instituto de F\'isica de Cantabria (CSIC-UC). Avda. Los Castros s/n. 39005 Santander, Spain.\and
	\small$^{37}$IPAC, California Institute of Technology, 1200 E. California Blvd. Pasadena, CA 91125, USA.\and
	\small$^{38}$Kavli Institute for Astronomy and Astrophysics, Peking University, Beijing 100871, China.\and
	\small$^{39}$Department of Physics and Astronomy, Rutgers, the State University of New Jersey, \and 
    \small Piscataway, New Jersey 08854, USA.\and
	\small$^{40}$Observat\'orio Nacional, Rua General Jos\'e Cristino, 77 - Bairro Imperial de S\~ao Crist\'ov\~ao, \and 
    \small Rio de Janeiro, 20921-400, Brazil.\and
	\small$^{41}$Laboratory for Multiwavelength Astrophysics, School of Physics and Astronomy, Rochester Institute of Technology, \and 
    \small 84 Lomb Memorial Drive, Rochester, NY 14623, USA.\and
	\small$^{42}$School of Physics \& Astronomy, University of Minnesota, 116 Church St. SE, Minneapolis, MN 55455.\and
	\small$^{43}$Center for Interdisciplinary Exploration and Research in Astrophysics (CIERA), \and 
    \small 1800 Sherman Ave., Evanston, IL 60201, USA.\and
	\small$^{44}$School of Physical and Chemical Sciences — Te Kura Matū, University of Canterbury, Private Bag 4800, \and 
    \small Christchurch 8140, Aotearoa, New Zealand.\and
	\small$^{45}$Centre for Astrophysics and Supercomputing, Swinburne University of Technology, Melbourne, VIC 3122, Australia.\and
	\small$^{46}$Observat\'orio do Valongo, Universidade Federal do Rio de Janeiro, \and 
    \small Ladeira do Pedro Antônio 43, Rio de Janeiro RJ 20080-090, Brazil.\and
	\small$^{47}$DTU-Space, Technical University of Denmark, Elektrovej 327, 2800, Kgs. Lyngby, Denmark.\and
	\small$^{48}$Institute of Science and Technology Austria (ISTA), Am Campus 1, 3400 Klosterneuburg, Austria.\and
	\small$^{49}$Department of Physics, Indian Institute of Science, Bengaluru 560012, India.\and
	\small$^{50}$MIT Kavli Institute for Astrophysics and Space Research, 70 Vassar Street, Cambridge, MA 02139, USA.\and
	\small$^{51}$Center for Frontier Science, Chiba University, 1-33 Yayoi-cho, Inage-ku, Chiba 263-8522, Japan.\and
	\small$^{52}$Department of Physics, Graduate School of Science, Chiba University, 1-33 Yayoi-Cho, Inage-Ku, \and 
    \small Chiba 263-8522, Japan.\and
	\small$^{53}$Department of Astronomy, University of Maryland, College Park, 20742, USA.\and
	\small$^{54}$D\'epartement d'Astronomie, Universit\'e de Gen\`eve, Chemin Pegasi 51, 1290 Versoix, Switzerland.\and
	\small$^{55}$Finnish Centre for Astronomy with ESO (FINCA), Quantum, Vesilinnantie 5, University of Turku, \and 
    \small FI-20014 Turku, Finland.\and
	\small$^{56}$Department of Physics, University of Helsinki, Gustaf Hällströmin katu 2, 00560 Helsinki, Finland.\and
	\small$^{57}$INAF – IASF Milano, via A. Corti 12, I-20133 Milano, Italy.\and
	\small$^{58}$Institute of Astronomy, Graduate School of Science, The University of Tokyo, 2-21-1 Osawa, \and 
    \small Mitaka, Tokyo 181-0015, Japan.\and
	\small$^{59}$INAF-OAS, Osservatorio di Astrofisica e Scienza dello Spazio di Bologna, via Gobetti 93/3, I-40129 Bologna, Italy.\and
	\small$^{60}$University of California, Santa Cruz, 1156 High St., Santa Cruz, CA 95064, USA.\and
	\small$^{61}$School of Earth and Space Exploration, Arizona State University, Tempe, AZ 85287-6004, USA.\and
	\small$^{62}$Department of Natural Sciences, The Open University of Israel, Ra'anana 4353701, Israel.\and
	\small$^{63}$Astrophysics Research Center of the Open University (ARCO), Ra'anana 4353701, Israel.\and
	\small$^\ast$Corresponding authors. Email: dcoulter@stsci.edu \& clarison@stsci.edu \\
	\small$^\dagger$These authors contributed equally to this work.
}

\begin{document} 

\maketitle


\begin{abstract} \bfseries \boldmath

We present the \textit{JWST} discovery of ``\snname'': a strongly lensed, multiply-imaged, SN II at a spectroscopic redshift of $\mathbf{z=5.133\pm0.001}$. SN~Eos exploded when the Universe was only $\mathbf{\sim1}$ billion years old, shortly after it reionized and became transparent to ultraviolet radiation. Archival \hst~imaging of \snname~reveals rest-frame far ultraviolet ($\mathbf{\sim1,300}$~\AA) emission, indicative of shock breakout or interaction with circumstellar material in the first few (rest-frame) days after explosion. Follow-up \textit{JWST} spectroscopy of \snname, now the farthest spectroscopically confirmed SN ever discovered, shows that \snname's progenitor star likely formed in a metal-poor environment ($\mathbf{\lesssim0.1~Z_{\odot}}$), providing strong, independent, and direct evidence for the chemical evolution of the Universe.
\end{abstract}

\clearpage

\noindent Beginning $400 - 500$ Myrs after the Big Bang, the neutral hydrogen that constituted the majority of the 
Universe underwent a phase transition driven primarily by ionizing UV radiation from small, intensely star-forming 
galaxies. This transition, which appears to have ended by $z\approx5.3$ \cite{bosman_reionization_2022}, is referred to 
as the Epoch of Reionization [EoR, \cite{becker_reionization_2001,Dayal2018_galaxy_formation}]. Direct observations of individual stars that existed during this early period would allow us to answer key questions about this phase in the Universe, such as how early stars formed the elemental building blocks of life, distributed these nucleosynthetic products to their surrounding ISM, and impacted their environment through stellar winds and ionizing radiation \cite{ceverino_feedback_2009}. Due to their extreme distances, the opportunities to study such stars remain quite limited \cite{windhorst_popiii_2018}. However, the explosive deaths of massive stars as core-collapse supernovae (CC~SNe), which can be brighter than the total emission of their host galaxies, allow us to probe the final stages of stellar evolution. CC~SN observations also reveal how heavy elements from stellar and SN nucleosynthesis, as well as unburnt elements that have been stored since the progenitor's birth, are ejected into the environment. These early SNe are likely to be formed in lower metallicity environments than in the local Universe \cite{nakajima_mass_metallicity_2023}, and thus could have stellar progenitors more representative of the first generation of stars than their local counterparts \cite{karlsson_2013}. In-depth studies of early-Universe SNe therefore provide crucial data to constrain early stellar evolution models, the interplay between SN properties and their host conditions, the cosmic star formation rate, and the potentially evolving initial mass function \cite{dessart_iip_2014}.

To detect these extremely distant objects, the Vast Exploration for Nascent, Unexplored Sources [VENUS, \cite{fujimoto_venus_2025}] \webb~treasury program targets clusters of galaxies that act as gravitational telescopes, dramatically boosting the apparent brightness of distant sources and therefore opening a window to SNe in the infant Universe \cite{natarajan_lensing_2024}. This ``strong'' lensing effect leads to the appearance of multiple magnified images of the background object, each with a different arrival time to the observer. The difference between these arrival times is known as the ``time delay'' and depends on the distances to the components of the lens system, thus directly probing cosmological evolution [e.g, the Hubble constant, \cite{refsdal_possibility_1964,treu_time_2016}]. The magnification effect, which can reach a factor of ten to hundreds -- or even thousands for rare sources near the thin regions of infinite magnification known as the ``critical curves'' \cite{diego_caustic_2018,williams_lensing_2024} -- provides the ability to detect distant sources, such as SNe and galaxies, that would have otherwise been too faint to discover. Detailed analyses of rare objects that leverage this increase in signal, such as lensed stars and explosive transients, can shed light on early-Universe physics \cite{welch_earendal_2022,diego_godzilla_2022,chen_shock_2022}.

\vspace{-0.5cm}

\section{Discovery of SN Eos} \label{sec:discovery}
\begin{figure}
    \centering
    \includegraphics[width=1.0\linewidth]{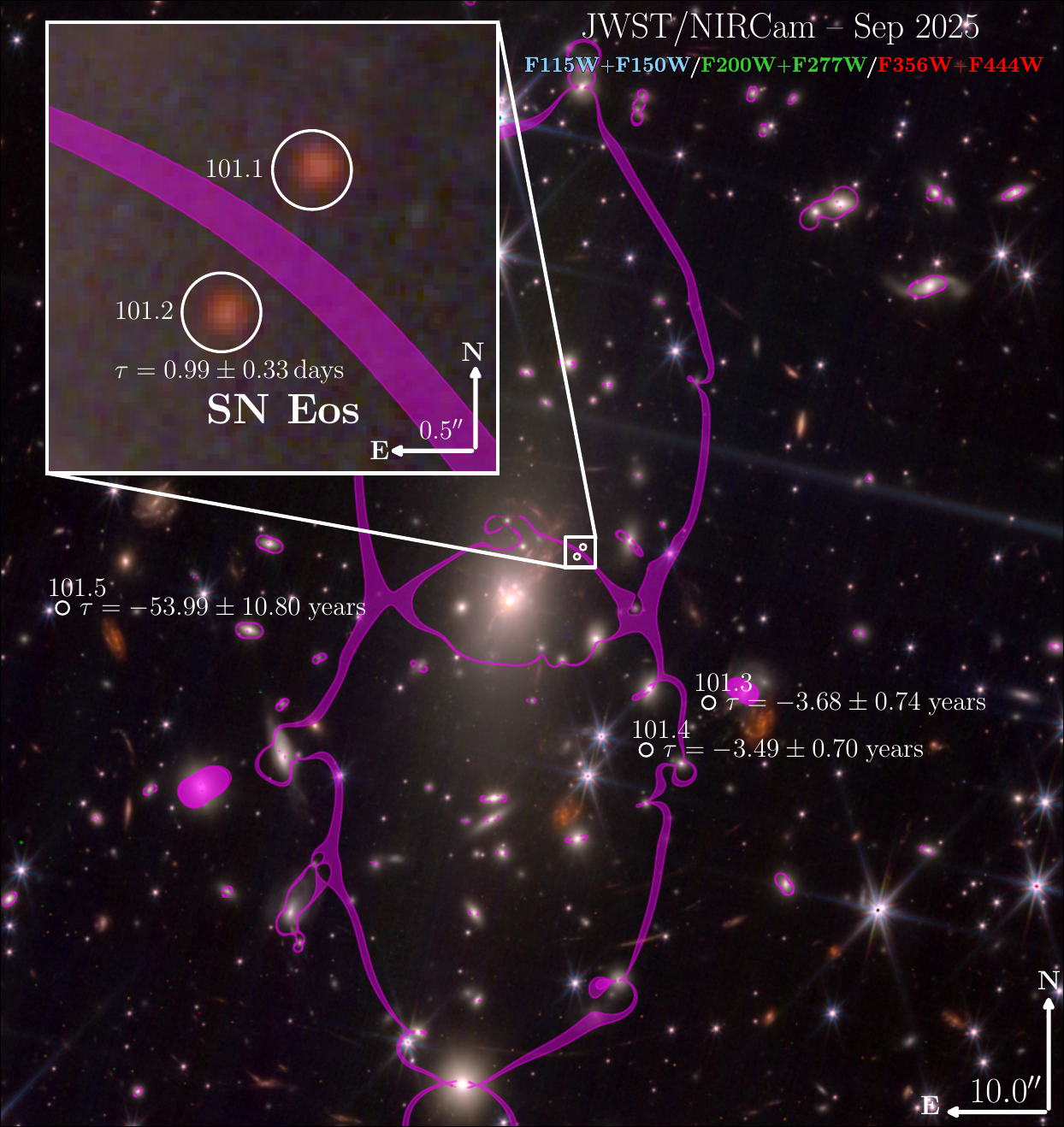}
    \caption{\small {\bf \textit{JWST} discovery image of the \clname\ galaxy cluster containing SN~Eos.} The RGB color channels are as follows: blue [F115W + F150W], green [F200W + F277W], and red [F356W + F444W]. The magenta shaded area corresponds to a region where the model-predicted magnification is $\mu > 100$ at $z=5.133\pm0.001$. The upper left inset shows the two detected images of \snname, 101.1 \& 101.2, mirrored by the critical curve, which represents the thin region of infinite magnification in the lens system for the source redshift. The proximity of these images to the critical curve results in a high magnification, predicted to be $\sim30$ at the image 101.1/101.2 positions. In the main image (outside the inset), we display the predicted positions and time delays, $\tau$, of images 101.3, 101.4, and 101.5. The angular scales are labeled on the N–E orientation arrows.}
    \label{fig:cluster}
\end{figure}

\noindent Here we present ``\snname'', a multiply-imaged, strongly lensed transient discovered as a pair of images observed on 2025 September 1 (all dates presented are UTC) by VENUS in \textit{JWST}/NIRCam imaging of the \clname\ [$z=0.35$; \cite{ciocan_muse_cluster_2021}] galaxy cluster field (see Figure~\ref{fig:cluster}). Follow-up NIRCam imaging and NIRSpec PRISM spectroscopy of \snname~taken on 2025 October 8 (\textit{JWST} DDT PID: 9493, see 
Figure~\ref{fig:spectral_plot}) reveal \snname~to be a hydrogen-rich, core-collapse supernova (SN II) based on the clear detection of the hydrogen Balmer P-Cygni profiles. After the initial \textit{JWST} discovery, an archival search of \textit{Hubble Space Telescope} ({\it HST}) imaging of the cluster field reveals a non-detection of \snname, or its host, at these locations 13 years prior in 2012 to a $5\sigma$ limiting magnitude of 27.3 AB mag in the WFC3/F814W and F110W filters. Repeat \hst~imaging of this field on 2024 March $11 - 18$ show $\sim10\sigma$ ($\sim25\sigma$) detections in F814W [F110W; see table~S1 in the Supplementary Text] on each of four (five) epochs over the course of this observer-frame week. At this redshift, the \hst~photometry corresponds to rest-frame far ultraviolet (FUV: $\sim1300-1900$ \AA). A further archival search of data at the location of \snname~from the Very Large Telescope (VLT) Multi Unit Spectroscopic Explorer (MUSE), taken in 2015 June [prior to the SN explosion, \cite{ciocan_muse_cluster_2021}], displays an isolated 
$\lambda7456$~\AA~emission line (rest-frame $1215$~\AA) associated with \lyalong~(\lya) from the underlying host galaxy (Figure~\ref{fig:host_LyA}). Fitting for the redshift with this line, we find $z_{spec}= 5.133 \pm 0.001$, which clearly aligns with the emission line peaks in the \snname~spectra.

\vspace{-0.5cm}

\section{Lens Model Description}
\noindent Our lens model for \clname, described in more detail in 
\cite{methods},
yields five predicted images for the \snname~+ host system (101.1-5). In two of these predicted images, 101.1-2, \snname~is detected in \textit{JWST}/NIRCam 
observations 
(see table~S1 in the Supplementary Text). 
In predicted images 101.3-4, strong \lya\ emission from the host galaxy was detected in pre-explosion MUSE data, while the fifth predicted image, 101.5, was not covered by the archival MUSE image footprint 
[see \cite{methods}]. 
As a result of the close proximity to the critical curve, images 101.1-2 are highly magnified (by factors of $\mu_{101.1}\approx25.46 \pm 5.73$ and $\mu_{101.2}\approx29.88 \pm 6.22$) and have a short, observer-frame predicted time delay of $\sim1$ day between them ($\sim4$~rest-frame hours), with image 101.2 being the last to arrive. 
\snname~was not detected in the other three locations; the model predicts that the light from \snname~arrived years earlier and was not as highly magnified. We summarize the magnification and time delays computed using the cluster lens model in 
table~S2 in the Supplementary Text, 
and throughout use a standard flat \LCDM\ cosmology with 
$H_0=70\,\mathrm{km}\,\mathrm{s}^{-1}\,\mathrm{Mpc}^{-1}$, $\Omega_{\Lambda}=0.7$, and $\Omega_{m}=0.3$.

\vspace{-0.5cm}

\section{Spectrophotometric Classification}
\begin{figure}
     \centering
     \includegraphics[width=1.0\linewidth]{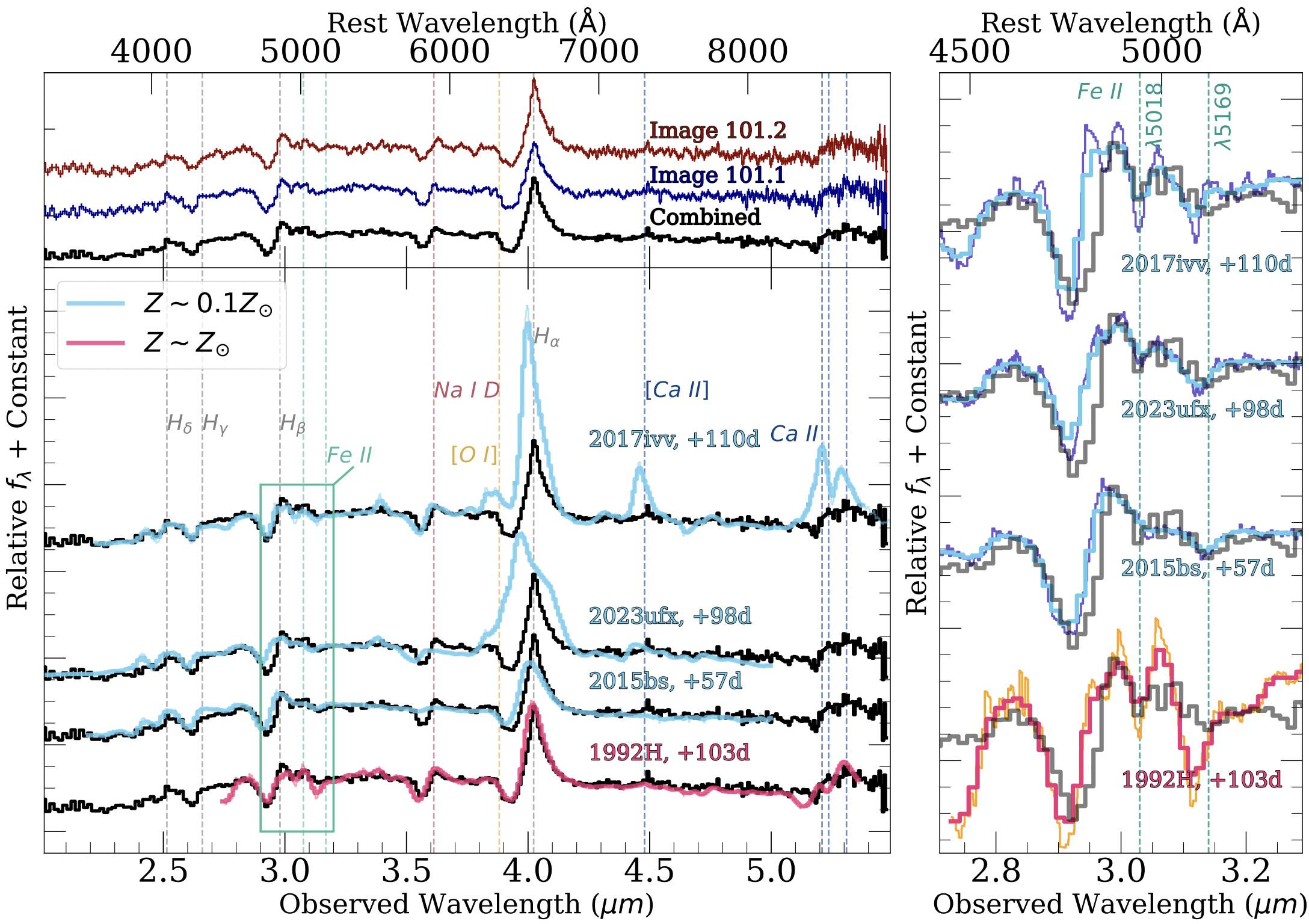}
     \caption{\small \textbf{Comparison of NIRSec PRISM spectra taken of \snname~(top-left) to local SNe~IIP (bottom-left) and comparisons of the Fe line profiles (right) of these same objects.}. {\it Top Left}: In red and blue, spectra of individual lensed images of \snname~are overlaid, demonstrating that each lens image shows the same SN at a similar phase. The inverse-variance weighted combined spectrum is shown in black and used in comparisons below. {\it Bottom Left}: Comparisons of our combined spectrum to the optical spectra of local SNe~IIP, normalized to their continua, convolved and resampled with the \webb~NIRSpec PRISM dispersion function. Spectra overlaid in blue are a sample of known, well-studied SNe~II with metallicities consistent with $Z \lesssim0.1~Z_{\odot}$: SN~2017ivv \cite{Gutierrez2020_2017ivv}, SN~2023ufx \cite{tucker_ufx_2024, Ravi2025_ufx}, and SN~2015bs \cite{anderson_2015bs_2018}. The overlaid spectrum in red corresponds to the best-matching local SN II, SN~1992H at $\sim1~Z_{\odot}$ 
     [see \cite{methods} for discussion of matching procedure]. 
     These comparisons demonstrate that SNe~IIP display many similarities despite having a range of metallicities, and even for SNe~IIP with similar metallicities there is still an intrinsic diversity in spectral features. This is consistent with findings from \cite{Anderson2016_SNeIIP_metallicity_probes} which claim that metallicity alone is not strongly correlated with all spectral observables. {\it Right}: A detailed view of the Fe~II complex for the same SNe, with \snname~in gray. Each local SN has its native spectral resolution (purple and orange) and NIRSpec PRISM-convolved resolution (blue and red) overplotted onto \snname.}
     \label{fig:spectral_plot}
\end{figure}

\begin{figure}
     \centering
     \includegraphics[width=1.0\linewidth]{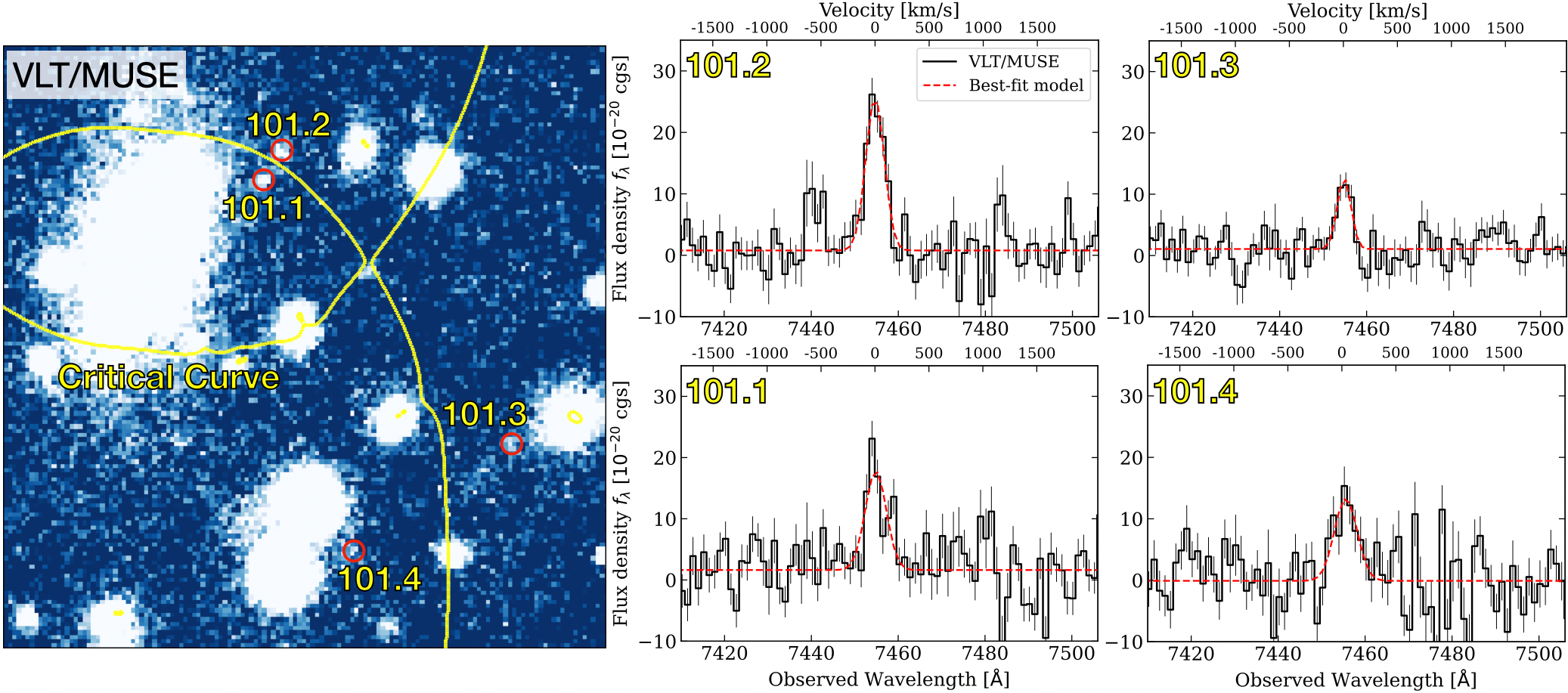}
     \caption{\small \textbf{VLT/MUSE observations of the SN Eos host galaxy.} {\it Left}: the moment-zero map centered on the narrow wavelength range of $\lambda_{\rm obs}=7452$ \AA\ to 7458 \AA, generated from MUSE data cube. Multiple images of the system (image 101.1, 101.2, 101.3, and 101.4) are detected. {\it Right}: VLT/MUSE spectra extracted at the positions of multiple images. The \lya~lines are detected at $\lambda_{\rm obs}=7455$ \AA\ in all images, with a measured redshift of $z= 5.133 \pm 0.001$. The best-fit single Gaussian profiles are shown in red.}
     \label{fig:host_LyA}
\end{figure}

\noindent Figure \ref{fig:spectral_plot} shows the spectroscopic observations of the two detected images (101.1 and 101.2) of \snname, taken 94~rest-frame days after the archival \hst~detections, which show a clear identification of the hydrogen Balmer lines, specifically \halpha, \hbeta, \hgamma, and \hdelta. These all show P-Cygni profile shapes that are characteristic of SNe, which consist of a blueshifted absorption component in addition to the emission line, and we unambiguously classify \snname~as a Type II supernova (SN II). We further note the presence of other common SN II elemental species, namely oxygen, sodium, and calcium. As shown in the top-left panel of Figure~\ref{fig:spectral_plot}, most of the commonly shared features between the spectra of the two images appear nearly identical, consistent with the small predicted time delay 
[see table~S2 in the Supplementary Text, as well as \cite{methods} for a detailed discussion].

We further classify \snname' SN subtype by considering its early UV light curve. In SNe~II, far-UV (FUV) radiation at the brightness we see in \snname~may be emitted within the first few days of explosion and is thus rarely detected. Generally, SNe~II FUV light curves rise within $\sim2-5$~days of explosion and then rapidly fade due to the SN ejecta expanding and cooling \cite{brown_sneiip_swift_2009,Pritchard2014_UV_LCs,bostroem_2022acko_2023}. The presence of strong FUV radiation in the archival \hst~imaging therefore points to the detection of \snname~within a few rest-frame days of explosion. Due to a lack of a constraining non-detection, however, we conservatively fix the explosion time, $t_{0}$, of \snname~to within five rest-frame days before the first \hst~detection. Combined with the redshift, this places our \webb~observations at approximate phases of $\sim92$ and $\sim98$ rest-frame days post-explosion. For SNe~II progenitors that retain massive hydrogen-rich envelopes, hydrogen recombination powers a relatively constant luminosity and results in a $\sim50 - 100$~rest-frame day ``plateau'' \cite{Anderson2014_SNIIP_LCs}, after which heating by the radioactive decay of $^{56}$Co and other long-lived radioactive elements becomes the primary source of luminosity. As further described in \cite{methods},
we compare our measured (demagnified) photometry to a grid of low-metallicity red supergiant (RSG) SN progenitor models, and conclude that \snname~is consistent with a SN~IIP at the end of its plateau phase -- as constrained by the UV detection.

\vspace{-0.5cm}

\section{Host Galaxy Analysis}
\noindent Apart from the aforementioned strong \lya~line (Figure \ref{fig:host_LyA}), the host of \snname~is faint and only marginally detected in the stacked rest-UV NIRCam images (F090W, F115W, and F150W), subtended by an angular length of $\sim0.4\arcsec$ and width of $\sim0.1\arcsec$. These detections lead to an apparent rest-UV magnitude of $28.5 \pm 0.4$~AB~mag for system 101.2, which corresponds to $M_{\rm UV} \approx -14.4$~AB~mag after correcting for magnification, line-of-sight Milky Way dust extinction \cite{schlegel_dust_1998,schlafly_dust_2011}, and applying the redshift-dependent \emph{K}-correction \cite{hogg_k_2002}. Given this combination, we identify the host of SN~Eos to be a \lyalong~emitter (LAE) galaxy, which are faint and metal-poor galaxies with high specific star formation rates at high-$z$ \cite{Saxena2024AAP}. According to the mass-metallicity relationship (MZR), fainter, lower-mass galaxies should have lower (gas-phase) metallicity. The fact that other high-redshift, low-mass galaxies and emitting regions observed with \textit{JWST} follow this relationship and are found to have $Z\lesssim0.1\ Z_\odot$ \cite{durkivica_lae_2025,morishita_lya_2025,vanzella_2025} provides supporting evidence that the gas-phase metallicity of the host of \snname, and thus the underlying environment of \snname, is also $<0.1\ Z_\odot$. Therefore, by leveraging strong lensing, our observations are probing the fainter (and lower metallicity) end of the LAE UV luminosity function compared to \textit{JWST} blank deep field surveys [e.g., \cite{Saxena2024AAP,Willott2025ApJ}]. Given the singular MUSE detection of the \lya~line at a low significance (and lack of any other identifying lines), this LAE would not have been identified without the presence of SN~Eos, suggesting that high-$z$ CC~SNe could help reveal a population of star-forming galaxies that are otherwise incredibly difficult to detect.

\vspace{-0.5cm}

\section{Low-z Comparison and Light Curve Modeling}
\noindent Given our extremely high signal-to-noise spectrum, \snname~provides a rare opportunity to directly test the inferred low metallicity of its LAE host. It has been predicted that because red supergiant (RSG) envelopes should be only weakly affected by nuclear burning, the presence of Fe-group elements in the hydrogen-rich envelope at late times is representative of the metallicity of the gas out of which the stellar progenitor formed  \cite{Dessart2011_SNIIP_models, Dessart2013_SNeIIP_properties, dessart_iip_2014}, thereby motivating their use as metallicity probes of the Universe. This prediction was supported by \cite{Anderson2016_SNeIIP_metallicity_probes, Taddia2016_SNIIP_metallicity, Gutierrez2018_SNIIP_metallicity} in the local Universe, finding a positive correlation with the strength of metal lines observed in SNe~IIP spectra and the metallicity inferred from local measurements of the host HII regions. By combining each lens image spectrum of \snname, we have the first opportunity to measure the ``pseudo-equivalent'' width (pEW) of Fe~II~$\lambda5018,5169$~\AA~in an SN~IIP near the EoR. We show in Figure~\ref{fig:spectral_plot} an overlay of strong- and weak-Fe features in local SNe~IIP, and notably contrast \snname' Fe II complex with known local, low-metallicity SNe 2015bs \cite{anderson_2015bs_2018}, SN~2017ivv \cite{Gutierrez2020_2017ivv}, and SN~2023ufx \cite{tucker_ufx_2024, Ravi2025_ufx}. This shows that the pEWs of \snname~are closely matched to these rare, $\lesssim0.1~Z_{\odot}$ examples.

\begin{figure}
     \centering
     \includegraphics[width=0.75\linewidth]{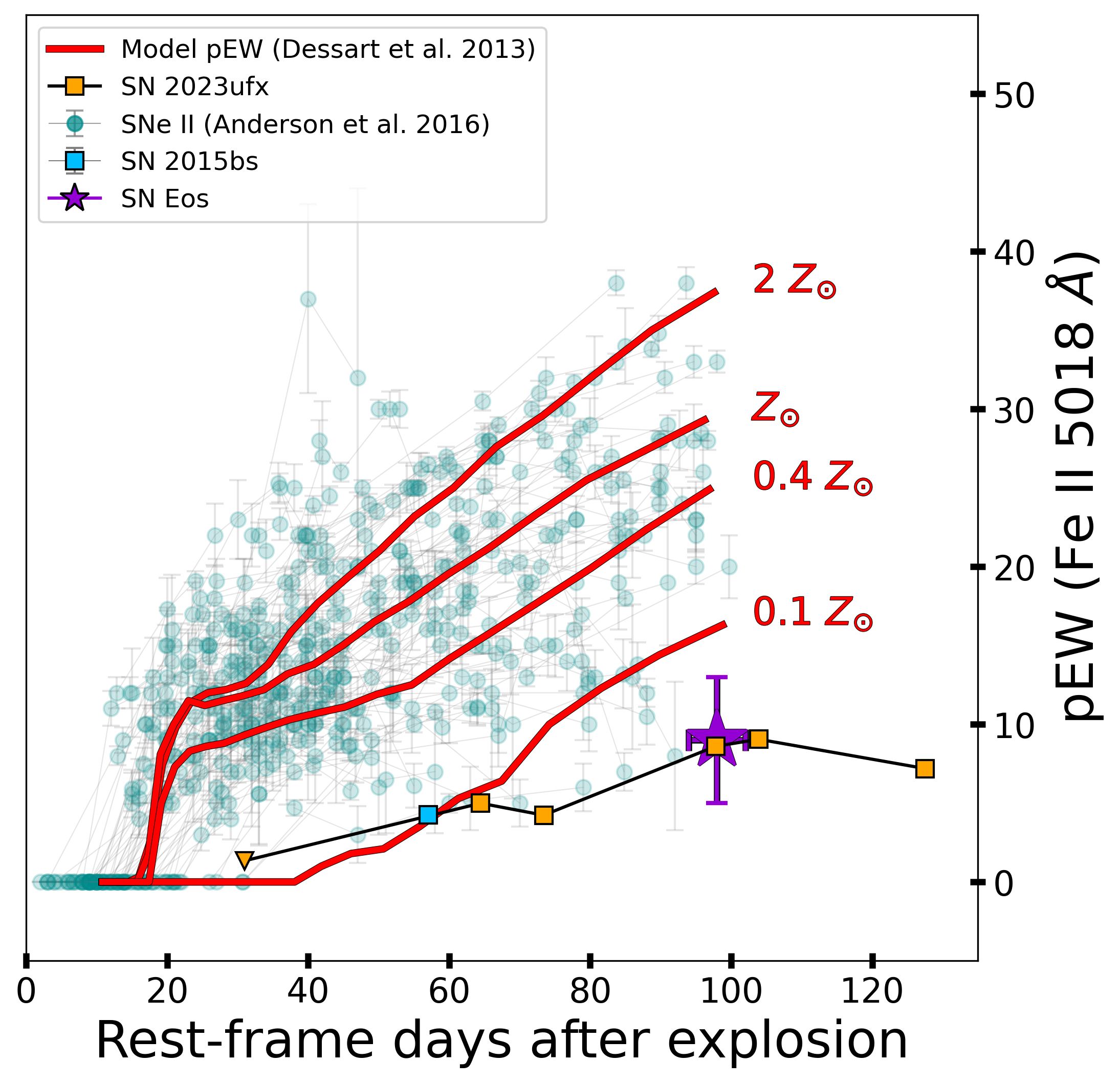}
     \caption{\small \textbf{The evolution of pEW Fe~II~$\mathbf{\lambda5018}$~\AA~measured in local SNe~IIP as a function of SN phase.} In teal are SNe data from \cite{Anderson2016_SNeIIP_metallicity_probes} with errors, and overlaid in red are spectral models as a function of metallicity from \cite{Dessart2013_SNeIIP_properties}. \snname~is shown as a purple star, with SN~2023ufx as orange squares, with the inverted triangle as a pEW upper limit \cite{tucker_ufx_2024}. A single measurement for SN~2015bs is plotted as a blue square \cite{anderson_2015bs_2018}. Our measurement, along with the inferred phase, places \snname in a likely parameter space that is $Z\lesssim0.1\ Z_\odot$.}
     \label{fig:eos_fe_II}
\end{figure}

We provide context for this measurement in Figure~\ref{fig:eos_fe_II} which follows the evolution of pEW Fe~II~$\lambda5018$~\AA~as a function of rest-frame SN phase. Overlaid are SNe~II data from \cite{Anderson2016_SNeIIP_metallicity_probes} along with spectral models as a function of metallicity from \cite{Dessart2013_SNeIIP_properties}. We caution that because our SN phase constraints from \hst~place our spectral observations at \spectralphase, \snname' photosphere would be cooler than most of those measured by \cite{Anderson2016_SNeIIP_metallicity_probes}, and therefore our measurement would appear deeper/higher metallicity than if it had been measured at an earlier phase. Therefore, we take our inferred metallicity from this measurement as an upper limit consistent with $Z\lesssim 0.1 Z_{\odot}$, in agreement with an order estimate of the metallicity expected from the ultra-faintness of the host and placing \snname~in the small class of known, extremely metal-poor SNe~IIP.

\begin{figure}
     \centering
     \includegraphics[width=1.0\linewidth]{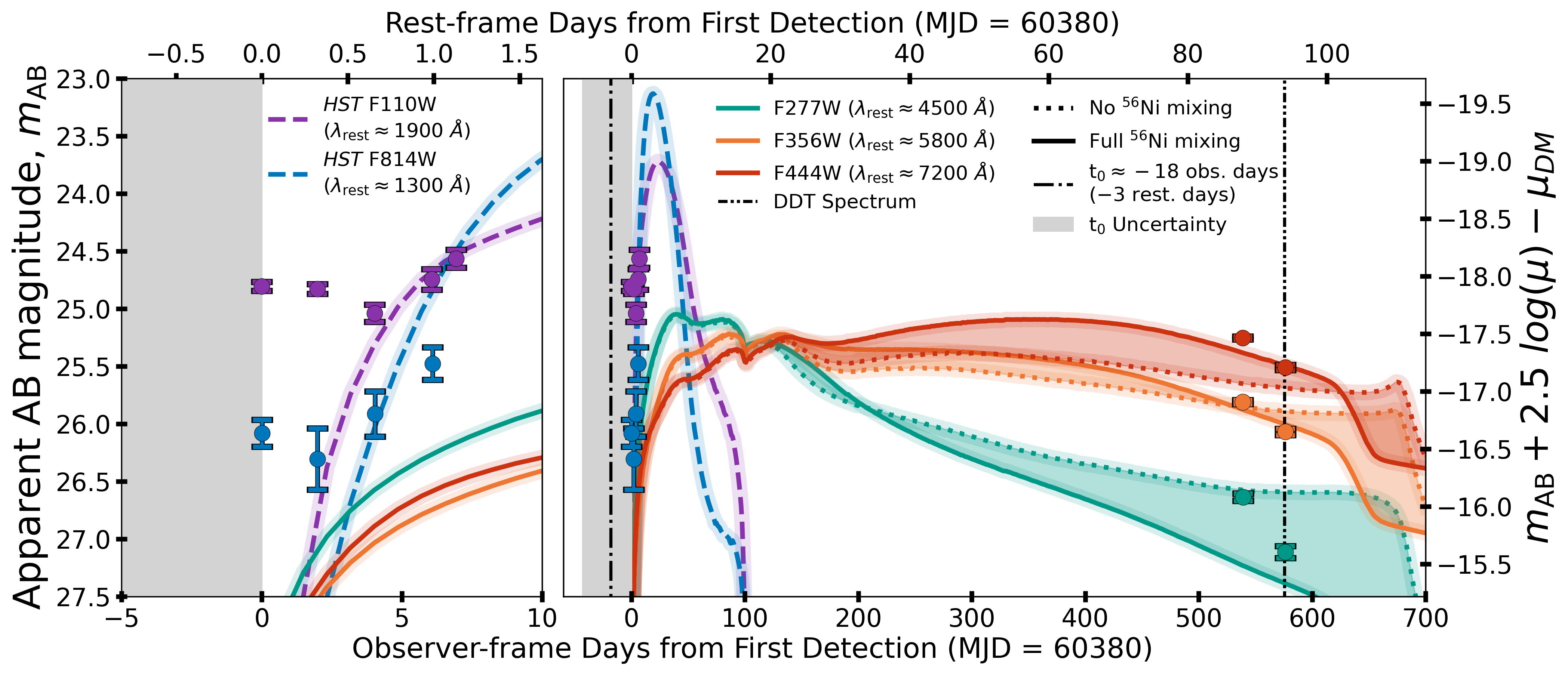}
     \caption{\small \textbf{Light curve of \snname, highlighting early FUV detections, with comparison against a theoretical model.} {\it Right}: \hst~and \webb~photometry for lens image 101.2, the higher signal-to-noise image, in case of a phase difference between 101.1 and 101.2 
     (see table~S2 in the Supplementary Text). 
     Overplotted are in-band light curves from the modeling in 
     \cite{methods},
     placing our \webb~observations at the end of an SN~IIP's plateau. The \hst~data were obtained as $4-5$ epochs over an observer-frame week, and are assumed to be near-explosion. The gray region marks the assumed window for $t_{0}$, the time of explosion, that we constrain to $\lesssim5$~rest-frame days from the first \hst~detection (see main text). {\it Left}: A zoomed-in view of the same \hst~F110W and F814W photometry, showing the variability of \snname~on a rest-frame $\sim1$~day timescale. The model prediction, containing a moderate amount of CSM material, is in rough agreement with the later luminosity and evolution of this UV emission, however, we note that the first $2$ epochs in F814W and F1101W are in excess of this simple CSM model suggesting that this flux could be due to more complicated effects, e.g., asymetric structure in the confined CSM 
     [see \cite{methods}].
     }
     \label{fig:lc_uv_jwst}
\end{figure}

We further model \snname~by fitting a series of self-consistent, hydrodynamical models to our \hst~and \webb~photometry. We note, however, that the model fit is highly sensitive to the absolute luminosity and light curve shape of \snname, and therefore to the lens model magnifications and time delays 
[see \cite{methods}].
To minimize biasing the fit by the uncertainties in the lens model, we select only the brighter lens image to fit ($101.2$, with $\mu \sim30\times$), calculate a distance modulus $\mu_{DM}\approx 46.4$~AB mag, and combine $\mu$ and $\mu_{DM}$ with several lens-independent observational constraints (e.g., the time of explosion, $t_{0}$, the apparent decline in the plateau, and inferred metallicity of $\lesssim 0.1~Z_{\odot}$) to produce the model fits displayed in Figure~\ref{fig:lc_uv_jwst}. Because it is possible that future lens models may produce differing magnifications for lens image $101.1$ and $101.2$, we simply present our observed photometry for lensed images $101.1$ and $101.2$ in 
table~S1 in the Supplementary Text.
We report that our best models are of a typical SN~IIP (in mass and explosion energy) with a slightly more luminous $B$-band peak than the average SN~IIP [$-17.7$~AB~mag vs. $\bar{M}_{B} = -16.80~\pm~0.37$~AB~mag; \cite{Richardson2014_AbsMag_dist}], with two different prescriptions for $^{56}$Ni mixing, and a spherical shell of confined CSM that relates the early rest-frame UV data to our rest-frame optical observations at $\sim +98$~days. In the left panel of Figure~\ref{fig:lc_uv_jwst} we zoom in on this early UV light curve and note that there is unaccounted for UV flux in the first two epochs that could be consistent with a more complex interaction scenario. Despite this, our simplified model is consistent with the subsequent UV rise and relative luminosity of the \hst~F110W and F814W photometry. A detailed description of this modeling process, our best-fit model, and our treatment of the UV light curve is presented in \cite{methods}.

As shown in Figure~\ref{fig:spectral_plot}, it is clear that \snname~shares several properties with both low- and high-metallicity SNe. \snname~best matches the spectral features of the solar-metallicity SN~1992H, especially in its \halpha~and Na I profiles. However, it is clear that \snname' Fe~II complex better matches that of SN~2015bs, SN~2017ivv, and SN~2023ufx, the only local SNe II with very low metallicity. 
In \cite{methods} 
we provide a more detailed comparison between \snname and these local SNe, but we note that \snname~also better matches the three low-$Z$ SNe in terms of \snname' inferred luminosity, host galaxy mass, and host metallicity. Surprisingly, even for SNe~II of similar, very low metallicities, there is a large diversity in their spectral and photometric properties \cite{Anderson2014_blueshift_emission_SNII, Anderson2014_SNIIP_LCs, Gutierrez2014_spectral_photo_diversity}. It has been speculated that this dispersion in properties arises from the detailed physics of each individual explosion, and that factors ranging from the size and structure of the progenitor star, its mass-loss history, rotation, and existence of a binary companion can have non-trivial and degenerate effects on SN observables; e.g., peak luminosities, plateau durations, or the strength and shape of spectral features \cite{Heger2003, Dessart2013_SNeIIP_properties, Sanyal2017}. With our relatively small data set, we cannot confidently infer the progenitor properties of \snname, and assessing whether or not low-redshift, low-$Z$ SNe are true analogs to their early-Universe counterparts is still an open question. Despite this, our combined spectrum of \snname~presents the first genuine test of these analogs, and based on how ``typical'' \snname~appears, also supports the results of \cite{Anderson2016_SNeIIP_metallicity_probes} --- that metallicity alone is not strongly correlated with any spectrophotometric SN property other than the pEW of Fe~II. A definitive answer to this question will require the discovery of many more high-redshift and local, low-$Z$ SNe, with well-sampled light curves and spectra. Such a sample would allow for self-consistently modeling the entire population as done by \cite{Anderson2014_SNIIP_LCs, Gutierrez2017_SNeII_Diversity, taddia_carnegie_2018, Martinez2022_CSP1}, providing statistically robust measurements of SN progenitor and explosion properties as a function of redshift and metallicity. Furthermore, calibrating these SNe~II as metallicity probes would allow for a constraint on the rate of chemical enrichment as a function of redshift and host galaxy properties -- especially for incredibly faint galaxies -- informing the evolution of the MZR \cite{tremonti_mzr_2004}. 

\vspace{-0.5cm}

\section{Implications for Future High-redshift Studies}
The discovery of \snname, a strongly-lensed, multiply-imaged, and highly-magnified SN~II at $z=5.133\pm0.001$ embedded in a very faint, LAE host galaxy, has allowed us for the first time to compare SNe in the local Universe to a SN close to the EoR. With an exquisite, high signal-to-noise follow-up NIRSpec spectrum rivaling the spectra taken of local SNe, we can unambiguously classify \snname~and probe the physics its explosion. Our spectrum reveals weak absorption of Fe~II~$\lambda5018,5169$~\AA, a known tracer of the gas-phase metallicity of the SN environment, and extends the use of SNe~II as metallicity tracers to the early Universe. Our measurement confirms that \snname~exploded in an environment with a metal concentration $\lesssim 10\%$ that of the solar abundance. In the local Universe, these low-metallicity SNe are incredibly rare ($<<1\%$ of all SNe~II); the fact that the {\it first} SN~II at $z>5$ with a high enough fidelity spectrum to directly assess its metallicity has such a low value is itself strong, independent, and direct evidence for the chemical evolution of the Universe. Serendipitous \hst~observations, prior to the \webb~discovery, show that in the rest-frame, \snname~had variable, bright, and rising FUV emission. We model this early UV light curve to emission occurring just days after explosion, and posit that a secondary rise in the UV is consistent with a moderate amount of confined, circumstellar material around the progenitor star. We compare \snname~to local examples of SNe~II, focusing on both the best, low-metallicity examples in the literature, as well as our best spectral match at solar-metallicity, and find that \snname~shares similarities with both sets of SNe. Finally, we emphasize that because SNe~II arise from deaths of massive stars, their progenitors have short lifetimes and trace the instantaneous star formation rates of their hosts. Integrating SN~II follow-up observations in high-$z$ galaxy surveys can therefore provide a more unbiased way of studying the faint end of the galaxy luminosity function compared with methods that assume a model \cite{Kim2023_Low_SB_galaxies}. Our results demonstrate that by leveraging the magnification afforded by gravitational lensing surveys, we have opened a new frontier in precision, high-redshift transient science. The discovery of \snname~represents a critical step toward fulfilling \webb's core mission objectives of understanding the lives and deaths of the first stars, the origins of the elements, and the assembly and evolution of the youngest galaxies.

%

\clearpage
\section*{Acknowledgments}

\paragraph*{Funding:}
We acknowledge the support of the Canadian Space Agency (CSA) [25JWGO4A18]. Numerical computations were in part carried out on a PC cluster at the Center for Computational Astrophysics, National Astronomical Observatory of Japan. This work is  based on observations made with the NASA/ESA/CSA James Webb Space Telescope. The data were obtained from the Mikulski Archive for Space Telescopes at the Space Telescope Science Institute, which is operated by the Association of Universities for Research in Astronomy, Inc., under NASA contract NAS 5-03127 for JWST. These observations are associated with JWST PID 6882. Support for PID 6882 was provided by NASA through a grant from the Space Telescope Science Institute. The authors acknowledge the use of the Canadian Advanced Network for Astronomy Research (CANFAR) Science Platform operated by the Canadian Astronomy Data Center (CADC) and the Digital Research Alliance of Canada (DRAC), with support from the National Research Council of Canada (NRC), the Canadian Space Agency (CSA), CANARIE, and the Canadian Foundation for Innovation (CFI).

C.L. and D.A.C. acknowledge funding through \textit{JWST} program grants JWST-GO-06541, JWST-GO-06585, and JWST-GO-05324. R.A. acknowledges support of Grant PID2023-147386NB-I00 funded by MICIU/AEI/10.13039/501100011033 and by ERDF/EU, and the Severo Ochoa award to the IAA-CSIC CEX2021-001131-S and from grant PID2022- 136598NB-C32 “Estallidos8”. H.A. acknowledges support from CNES, focused on the \textit{JWST} mission, and the French National Research Agency (ANR) under grant ANR-21-CE31-0838. F.E.B. acknowledges support from ANID-Chile BASAL CATA FB210003, FONDECYT Regular 1241005, and ECOS-ANID ECOS240037. M.B. acknowledges support from the ERC Grant FIRSTLIGHT, Slovenian national research agency ARIS through grants N1-0238 and P1-0188, and ESA PRODEX Experiment Arrangement No. 4000149972. M.B. and T.P. acknowledge the financial support from the Slovenian Research Agency (grants I0-0033, P1-0031, J1-8136, J1-2460, Z1-1853 and N1-0344) and the Young Researchers program, as well as the financial support from European Space Agency PRODEX Experiment Arrangement EArTH. C.J.C. acknowledges support from the ERC Advanced Investigator Grant EPOCHS (788113). P.D. warmly acknowledges support from an NSERC discovery grant (RGPIN-2025-06182). S.D. is supported by UK Research and Innovation (UKRI) under the UK government’s Horizon Europe funding Guarantee EP/Z000475/1. J.M.D. acknowledges the support of projects PID2022-138896NB-C51 (MCIU/AEI/MINECO/FEDER, UE) Ministerio de Ciencia, Investigaci\'on y Universidades and SA101P24. P.L.K. acknowledges funding from STScI GO-17504 and NSF AAG 2308051. I.L. acknowledges support by the Australian Research Council through Future Fellowship FT220100798. Z.G.L. is supported by the Marsden Fund administered by the Royal Society of New Zealand, Te Ap\={a}rangi grants M1255. X.L. is supported by an LSST-DA Catalyst Fellowship through the support of Grant 62192 from the John Templeton Foundation to LSST-DA. G.E.M. acknowledges the Villum Fonden research grants 37440 and 13160. N.M. acknowledges support from the ERC Grant FIRSTLIGHT and the Slovenian National Research Agency ARRS through grant N1-0238. G.N. acknowledges support by the Canadian Space Agency under a contract with NRC Herzberg Astronomy and Astrophysics. M.P. is supported by NASA through an Einstein Fellowship grant No. HST-HF2-51583.001-A awarded by the Space Telescope Science Institute (STScI), which is operated by the Association of Universities for Research in Astronomy, Inc., for NASA, under contract NAS5-26555. T.P. acknowledges the financial support for the bilateral collaboration between STScI and UNG (BI-US/24-26-085). M.R.S. is supported by an STScI Postdoctoral Fellowship. Q.W. is supported by the Sagol Weizmann-MIT Bridge Program. J.R.W. acknowledges that support for this work was provided by The Brinson Foundation through a Brinson Prize Fellowship grant. R.A.W. acknowledges support from NASA JWST Interdisciplinary Scientist grants NAG5-12460, NNX14AN10G and 80NSSC18K0200 from GSFC. Y.X. is supported by JSPS KAKENHI Grant Number JP25KJ1029. A.Z. acknowledges support by the Israel Science Foundation Grant No. 864/23. 

\paragraph*{Author contributions:}
D.A.C. and C.L. led the analysis of the main text and supplementary material, contributing to each figure and table. J.D.R.P. worked on the photometry of SN~Eos, including work on Figures 5, S1 and Table S1. S.F. is the PI of VENUS and led coordination of analysis efforts for this work. V.K. led the data reduction process for the \textit{JWST} and \textit[HST] data used in this paper. J.F.V.A. constructed the lens model used in this paper, as well as contributed to the analysis of the MUSE dataset. T.J.M. worked on the light curve modelling of SN~Eos, directly contributing to Figure~5. M.S. helped with the spectroscopic reduction of the \textit{JWST} NIRSpec observations and contributed extensively to Figure 2. Y.A. contributed analysis of the host galaxy of SN~Eos and contributed to Figure 3. R.B. and M.B. provided insight into the interpretation of the lens model and physical nature of SN Eos. G.B. helped with both data reduction and analysis. J.C., D.C., P.D., M.E., S.L.F., O.D.F., L.J.F., A.M.K., T.M., M.N., M.O., R.P., R.Q., A.R., J.R., L.R., L.G.S., F.S., T.T., H.Y. are key members of the VENUS team who provided valuable insight into the SN and lens modeling analysis. A., A.A., R.A., J.P.A., R.A., H.A., F.E.B., L.D.B., V.B., M.B., C.J.C., C.D., J.M.D., G.D., S.D., J.M.D., E.E., A.F., B.F., S.G., M.G.-O., M.G., Y.H., K.I., S.W.J., Y.J.-T., J.S.K., P.L.K., L.A.K., Z.G.L., X.L., I.L., P.A.A.L., R.A.L., G.E.M., N.S.M., J.M., A.K.M., R.P.N., G.N., M.O., E.P.G., M.P., T.P., M.R., D.S., S.S., M.S., W.S., K.S., A.T., E.V., Q.W., J.W., R.W., R.W., Y.X., Y.Z., A.Z. provided valuable analysis into the lens modeling and SN physics interpretations of this article.

\paragraph*{Competing interests:}
``There are no competing interests to declare.''

\paragraph*{Data and materials availability:}
All \textit{HST} and \textit{JWST} data presented and analyzed in this article can be accessed here: \url{https://www.doi.org/10.17909/r2dr-mq75}. The MUSE data can be accessed via the European Southen Observatory archive: \url{https://tinyurl.com/muse-eos-data}.

\clearpage
\input{supp_materials}

\bibliography{science_template,references} 
\bibliographystyle{sciencemag}

\end{document}

%% file: supp_materials.tex
\linespread{1.5} 

\frenchspacing

\renewenvironment{abstract}
	{\quotation}
	{\endquotation}

\date{}

\renewcommand\refname{References and Notes}

\makeatletter
\renewcommand{\fnum@figure}{\textbf{Figure \thefigure}}
\renewcommand{\fnum@table}{\textbf{Table \thetable}}
\makeatother

\renewcommand{\thefigure}{S\arabic{figure}}
\renewcommand{\thetable}{S\arabic{table}}
\renewcommand{\theequation}{S\arabic{equation}}
\renewcommand{\thepage}{S\arabic{page}}
\setcounter{figure}{0}
\setcounter{table}{0}
\setcounter{equation}{0}
\setcounter{page}{1} 



\begin{center}
\section*{Supplementary Materials for\\ A spectroscopically confirmed, strongly lensed, metal-poor Type II supernova at z = 5.13}

David A. Coulter$^{*\dagger}$,
Conor Larison$^{*\dagger}$,
Justin D. R. Pierel,
Seiji Fujimoto,
Vasily Kokorev,
Joseph F. V. Allingham,
Takashi J. Moriya,
Matthew Siebert,
Yoshihisa Asada,
Rachel Bezanson,
Maru\v{s}a Brada\v{c},
Gabriel Brammer,
John Chisholm,
Dan Coe,
Pratika Dayal,
Michael Engesser,
Steven L. Finkelstein,
Ori D. Fox,
Lukas J. Furtak,
Anton M. Koekemoer,
Thomas Moore,
Minami Nakane,
Masami Ouchi,
Richard Pan,
Robert Quimby,
Armin Rest,
Johan Richard,
Luke Robbins,
Louis-Gregory Strolger,
Fengwu Sun,
Tommaso Treu,
Hiroto Yanagisawa,
Abdurro'uf,
Aadya Agrawal,
Ricardo Amor\'{i}n,
Joseph P. Anderson,
Rodrigo Angulo,
Hakim Atek,
Franz E. Bauer,
Larry D. Bradley,
Volker Bromm,
Mateusz Bronikowski,
Christopher J. Conselice,
Christa DeCoursey,
James M. DerKacy,
Guillaume Desprez,
Suhail Dhawan,
Jose M. Diego,
Eiichi Egami,
Andreas Faisst,
Brenda Frye,
Sebastian Gomez,
Mauro Gonz\'alez-Otero,
Massimo Griggio,
Yuichi Harikane,
Kohei Inayoshi,
Saurabh W. Jha,
Yolanda Jim\'enez-Teja,
Jeyhan S. Kartaltepe,
Patrick L. Kelly,
Lindsey A. Kwok,
Zachary G. Lane,
Xiaolong Li,
Ivo Lobbe,
Paulo A. A. Lopes,
Ray A. Lucas,
Georgios E. Magdis,
Nicholas S. Martis,
Jorryt Matthee,
Ashish K. Meena,
Rohan P. Naidu,
Ga\"{e}l Noirot,
Masamune Oguri,
Estefania Padilla Gonzalez,
Massimo Pascale,
Tanja Petrushevska,
Massimo Ricotti,
Daniel Schaerer,
Stefan Schuldt,
Melissa Shahbandeh,
William Sheu,
Koji Shukawa,
Akiyoshi Tsujita,
Eros Vanzella,
Qinan Wang,
John Weaver,
Robert Williams,
Rogier Windhorst,
Yi Xu,
Yossef Zenati,
Adi Zitrin

\small$^\ast$Corresponding authors. Email: dcoulter@stsci.edu \& clarison@stsci.edu \\
\small$^\dagger$These authors contributed equally to this work.

\end{center}

\subsubsection*{This PDF file includes:}
Materials and Methods\\
Figure S1\\
Tables S1 to S2\\

\newpage


\subsection*{Materials and Methods}\label{sec:methods}

\subsection{\textit{JWST} Discovery} \label{sec:jwst_data}

SN~Eos was discovered in \textit{JWST} NIRCam imaging acquired on 2025 September 1 (all dates presented will be UTC), by the Vast Exploration for Nascent, Unexplored Sources collaboration [VENUS, \cite{fujimoto_venus_2025}] in the \clname\ [$z=0.35$, \cite{ciocan_muse_cluster_2021}] galaxy cluster field. Observations of the cluster included 10 NIRCam filters: [short-wavelength (SW): F090W, F115W, F150W, F200W, F210M] and [long-wavelength (LW): F277W, F300M, F356W, F410M, F444W] and are part of the larger VENUS program to image $\sim60$ rich galaxy clusters with \textit{JWST}. The VENUS image reduction process is as follows. We begin with \textit{JWST} Level-2 MAST products and process them using the \texttt{grizli} pipeline \cite{brammer_grizli_2023}, following the approach used in the public DJA reduction\footnote{\url{https://dawn-cph.github.io/dja/}}. We use the CRDS context map \texttt{jwst\_1456.pmap} for the photometric calibration. Compared to the standard STScI pipeline, \texttt{grizli} applies enhanced procedures to alleviate effects from cosmic rays, scattered light and detector artifacts \cite{bradley_highz_cands_2023,rigby_jwst_perf_2023}. We further perform additional background, 1/f noise and diffraction spike subtraction at both the amplifier and mosaic levels [e.g., \cite{endsley_jwst_2023,kokorev_jwst_2025}]. Final NIRCam mosaics are then drizzled to native pixel scales for each detector (SW: $0.03\arcsec$/pix, LW: $0.06\arcsec$/pix).

\subsection{\textit{HST} Archival Detection}
\label{sec:hst}

\textit{Hubble Space Telescope} ({\it HST}) imaging of the cluster field reveals a clear early rest-frame far ultra-violet (FUV; $\sim1300-1900$ \textup{\AA}) detection of \snname. This imaging was acquired on 2024 March 11, 13, 15, 17, 18 as part of an \textit{HST} SNAP survey program to identify lensed stars [PID: 17504, \cite{kelly_snap_2023}] in the WFC3/UVIS F814W and WFC3/IR F110W filters. The total exposure times for these two filters on each day were, in chronological order and for F814W \& F110W respectively, [(2220 s, 1068 s), (1110 s, 609 s), (1110 s, 459 s), (880 s, 369 s), (880 s, 369 s)]\footnote{The fifth epoch of the WFC3/UVIS F814W data appears to have suffered a guide-star failure, and thus we do not include it in the analysis.}. In a similar manner as for \textit{JWST}, we process the \textit{HST} data from the level 2 MAST products. For F814W, we use the products that include the pixel-based corrections for charge transfer efficiency (CTE). For both filters and all epochs (table \ref{tab:phot}), we again drizzle exposures using the \texttt{grizli} pipeline \cite{brammer_grizli_2023}, following the approach used in the public DJA reduction. We use the native pixel scales for these products as well, corresponding to $0.04\arcsec$/pix for F814W and $0.13\arcsec$/pix for F110W. 

\subsection{VLT MUSE Archival Detection} \label{subsec:MUSE_Ly-alpha}

Further archival Very Large Telescope (VLT) Multi Unit Spectroscopic Explorer (MUSE) data was retrieved from the ESO data archive that was taken on 2015 June 12 and July 17 -- as a series of six exposures, each of 1462 s -- thus preceding the \textit{JWST} observation of images 101.1-2 (see Main Text Figure~1)
of \snname~by over a decade, or around 20 rest-frame months [data and analysis presented in \cite{ciocan_muse_cluster_2021}]. The locations of the SN appear coincident with the detection of an isolated emission line that corresponds to \lya~emission and was previously unreported. These detections are at a significance of 7.9\,$\sigma$, 12.0\,$\sigma$, 6.4\,$\sigma$, and 5.8\,$\sigma$ for images 101.1-4 respectively. Due to these data being taken well before the SN exploded in these images, this line is associated with the host galaxy of \snname~and is also seen at two of the other host galaxy image locations (shown in Main Text Figure~3)
. A further dedicated analysis on the properties of this host will be presented in Y. Asada et al. ({\it in prep.}).

\subsection{\textit{JWST} DDT Observations}
\label{sec:JWST_follow_up}

Spectrophotometric observations with \textit{JWST} were executed on 2025 October 8 (DDT PID: 9493, PI: Coulter). This included a NIRSpec PRISM/CLEAR spectroscopic observation of both visible images 101.1 and 101.2 (with 10,504 s of exposure each), as well as follow-up NIRCam imaging (with 805 s of exposure) in the F277W, F356W, and F444W bands, each paired with the F070W SW filter (in an attempt to detect the underlying host). The NIRCam imaging was reduced using the standard VENUS procedure outlined in Supplementary Text section~\ref{sec:jwst_data}, while the NIRSpec spectra were reduced using the tools included in the \texttt{msaexp} package \cite{brammer_msaexp_2023}.

\subsection{Photometry}
\label{sec:phot}
\begin{figure}[h!]
    \centering
    \includegraphics[width=\linewidth]{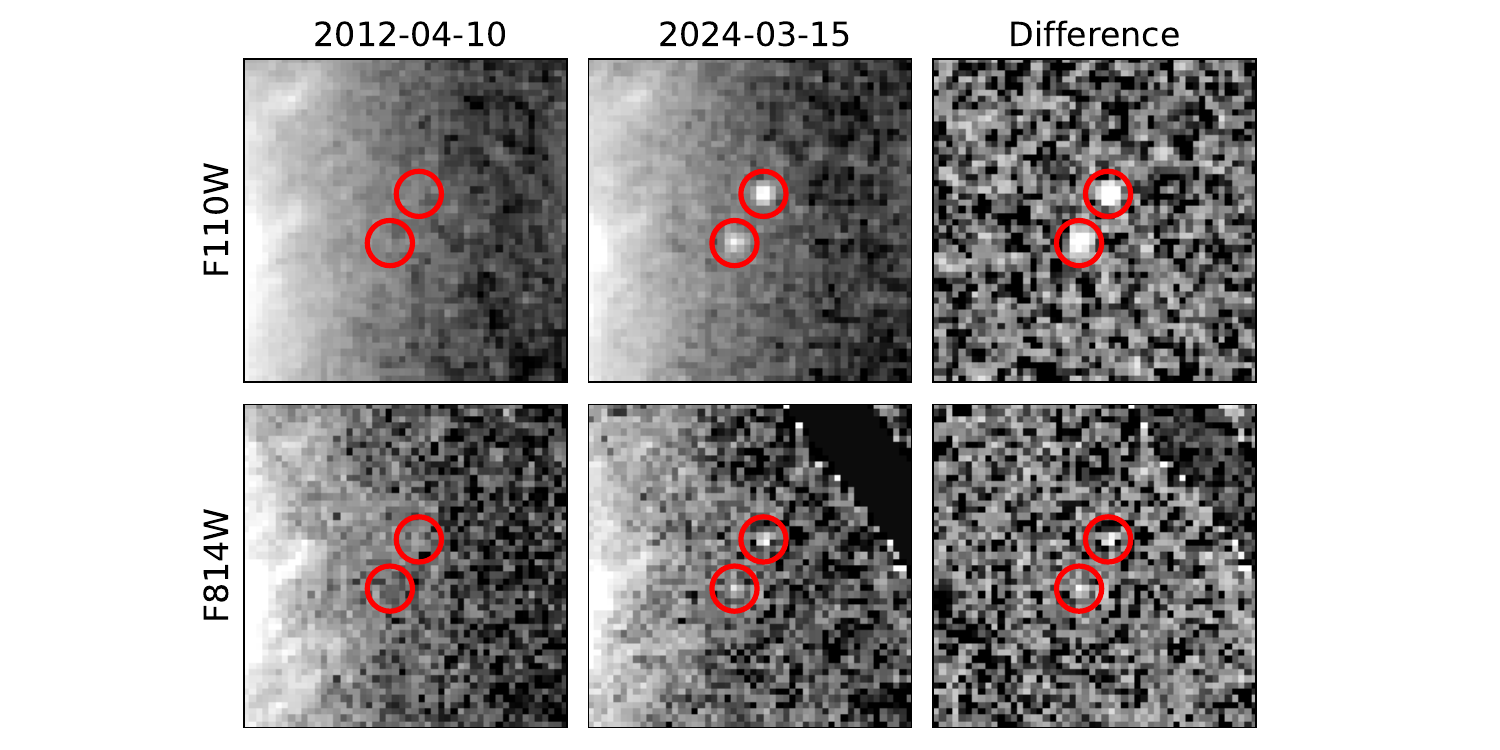}
    \caption{\small \textbf{The archival detection of \snname~in \textit{HST}~imaging, described in Supplementary Text section \ref{sec:hst}.} The top row shows the F110W filter, and the bottom row shows the F814W filter. Template imaging from 2012 (left column) shows no point source, while subsequent imaging from 2024 has clear detections of both SN images. The right column shows the difference between the 2024 and 2012 observations.}
    \label{fig:hst_detection}
\end{figure}

We use \texttt{STARRED} \cite{Michalewicz2023, Millon_2024} to measure \textit{HST} photometry of \snname~using the final drizzled products described in Supplementary Text section \ref{sec:jwst_data}, which is a Python package for scene modeling PSF photometry, optimized for light curve extraction.  In \texttt{STARRED}, the background is modeled on a grid of pixels with half the pixel size of the original image and regularized using starlets \cite{starck_starlet_2015}. We use the template imaging (figure \ref{fig:hst_detection}) to constrain this static background, and then \texttt{STARRED} takes as input all time-series images along with a PSF model for each epoch to measure the SN light curve. The PSF is reconstructed with \texttt{STARRED} for each filter and epoch using 8 non-saturated stars in the field of view, and then each \textit{HST} band is processed independently. 

\texttt{STARRED} does not perform well when there is no template imaging, as there is a degeneracy between the background reconstructon and the point source flux \cite{pierel_cosmology_2025}. Therefore for \textit{JWST}, where no template exists, we rely on robust aperture photometry with $3$ pixel radii, implemented on the final drizzled products, and background estimation using an annulus with inner and outer radii defined by the default reported aperture corrections (also applied to the raw measured fluxes) for each instrument. As all \textit{HST} and \textit{JWST} images were converted to the same units ($10$\,nJy) by the \texttt{grizli} pipeline, all flux measurements were simply converted to AB magnitudes directly with the corresponding zero-point of $28.9$. All measured photometry, including $5\,\sigma$ upper-limits, are reported in table \ref{tab:phot}.

\subsection{Lens Model} \label{sec:lens_model}

We construct a lens model for the galaxy cluster \clname\ using a Python-implemented, updated version of the parametric method by [\cite{Zitrin2015}, but see also \cite{pascale22, Furtak2022UNCOVER} where more details are provided]. The lens model and its implementation will be detailed in separate companion article (Allingham et al., in prep.), and here we summarize its main properties.

The lens model consists of a dark-matter component and a galaxy component.
With the cluster appearing mostly relaxed, we model its dark-matter component with one large, smooth dark-matter halo, corresponding to a pseudo-isothermal elliptical mass distribution [PIEMD; \cite{Keeton2001models}]. Individual cluster galaxies are selected via a multi-band selected red-sequence, and are modeled each using a dual pseudo-isothermal elliptical mass distribution [dPIE, see \cite{Eliasdottir2007arXiv0710.5636E}].
The cluster galaxies are scaled according to the Faber-Jackson relation \cite{FaberJackson}, but the brightest cluster galaxy (BCG), and its ellipticity, are allowed to be modified independently.

Our lens model is based on 51 multiple images of 19 background sources that are used as constraints, 7 of which were previously known, e.g.\ from the CLASH survey \cite{PostmanCLASH2012,Zitrin2015}, and 12 are newly identified with NIRCam in the VENUS program (including the images of the SN and the host). The parameter values are optimized via a $\chi^2$ function minimizing the distance between observed images and their predicted locations. The parameter space exploration was performed through an MCMC with stochastic moves guided by relative likelihood gradients.

The final best-fit model, minimized using 40 constraints and 21 free parameters, presents an $r.m.s.$ error between the model predictions and observations of $0.44''$. Owing to stringent constraints requested on the position of \snname, the two existing images are reproduced with a precision lower than the pixel resolution ($0.04''$). Due to the low input positional uncertainties, the resultant time delay and magnification uncertainties are most likely underestimated. Thus, we adopt a nominal 20\% additional systematic uncertainty on both throughout. The derived magnifications and time delays, located at the model-predicted image positions, are presented in table\,\ref{tab:detections}.

\subsection{Fe II pEW Measurement} \label{sec:fe_measurement}

To measure the pseudo-equivalent width pEW of Fe II $\lambda$ 5018, we first smooth the spectrum using the inverse-variance 
Gaussian algorithm from \cite{branch06}. From this smoothed spectrum, we identify the local maxima in flux on either side of the
absorption feature to act as the endpoints of the linear pseudo-continuum. The value of the pEW is determined within this region and the uncertainty is estimated via a simple Monte Carlo simulation, drawing simulated fluxes from the flux uncertainties at each wavelength, and varying the smoothing scale for the continuum definition. We find a Fe II $\lambda 5018$ pEW = $9 \pm 4~\mbox{\AA}$, a $2.3\,\sigma$ detection (see Main Text Figure~2 and Figure~4)
. We caution that given the low resolution of the detector, blending of adjacent SN features may bias this measurement. We also caution that the phase of SN Eos where we make this measurement is very near the end of the plateau, where the ionization state is changing due to a drop in the characteristic plateau temperature of $\sim5000$~K \cite{Martinez2022_CSP1}. Nonetheless, this pEW is broadly consistent with an O3N2-derived metallicity of $\lesssim0.1$~Z$_{\odot}$ when considering that the pEW monotonically increases with phase, and is consistent with the expectation given the host luminosity.

Furthermore, we apply the same methods to measure the pEW of Fe II $\lambda$ 5169. This feature is more strongly detected with Fe II $\lambda 5169$ pEW = $20 \pm 3~\mbox{\AA}$. While there may be biases associated with deriving these values from low-resolution spectra, and we cannot rule out a potential contribution from the host galaxy [O III], we note that the Fe II absorption complex appears quite similar in morphology to the other low-metallicity local SNe-II (blue curves, Main Text Figure~4).

\subsection{Spectroscopic Matching} \label{sec:spectral_diffs}

We compare the two spectra of \snname~to an extended library of low-redshift SN templates \cite{2025RNAAS...9...78M} using \texttt{SNID-SAGE} \cite{snid_sage_2025}, a Python implementation of the original Supernova Identification [SNID, \cite{2007ApJ...666.1024B}] code. Both spectra matched best to the +112 and +103 day spectra of SN~1992H, a Type IIP (plateau) SN that occurred in NGC 5377. Further shared matches include the low-metallicity Type II SN~2015bs, and the Type IIP SN~2013fs.

To compare the spectra of the two images of \snname~directly to one another, we first scale the brighter 101.2 spectrum to the 101.1 spectrum using a Markov Chain Monte Carlo (MCMC) fit assuming a Gaussian posterior and uninformed prior over the spectral range of 2.2 -- 5.5 $\mu$m. This gives a scale factor of $0.95\pm0.05$, consistent with the lens model-predicted value of $0.85\pm0.26$. Since the continuum emission is much brighter than the contribution from individual lines, this process yields consistent results when fitting to regions of the spectra that lack emission and absorption features. As shown in Main Text Figure~2
, most of the commonly shared features between the spectra of the two images appear identical. We do note, however, that there appears to be a significant deviation in the flux of the \halpha~emission feature after scaling. Integrating the spectral flux density of the two spectra from 3.97 to 4.13 $\mu$m, propagating the scale factor uncertainty into the significance, we find that the \halpha~flux is $\sim6~\sigma$ brighter for image 101.1 than image 101.2. If we instead scale the two spectra based on the observed photometry, we can derive a scale factor of $0.874\pm0.005$, in $2~\sigma$ agreement with the value derived directly from the spectra and in strong agreement with the lens model prediction. Using this scale factor instead, the significance in the \halpha~difference is on the order of $13~\sigma$. We believe that this difference in \halpha~strength is most likely resulting from contamination by the host galaxy. Due to the host being resolved, the total magnification factor affecting each image is different. This difference in magnification thus results in unique amounts of host contamination in the \halpha~line for each image, resulting in inconsistent \halpha~line strengths of the combined SN + host spectra. Based on the agreement in all the other line features, we believe that only \halpha~is significantly contaminated by the host, and we find no other evidence that could point to an alternate explanation, such as a large time delay or reduction artifact. An in-depth discussion of the underlying host contamination will be presented in Y. Asada et al. ({\it in prep.}).

\subsection{SN Eos Modeling} \label{sec:eos_modeling}
\label{sec:modeling}

To estimate the explosion properties of \snname, we compare synthetic light curves with the photometry of SN Eos presented in table~\ref{tab:phot}, demagnified by a factor of $\mu=30\times$. We followed the same procedures as our previous studies \cite{coulter_discovery_2025,moriya_jades_2025} on high-redshift Type~II SNe to obtain synthetic light curves. We took the red supergiant (RSG) SN progenitor models in \cite{moriya_jades_2025} having 0.1 Z$_{\odot}$, in agreement with both the metallicity inferred from the analysis of the LAE host and directly from SN Eos itself. The RSG progenitors were computed by the stellar evolution code \texttt{Modules for Experiments in Stellar Astrophysics (MESA)} \cite{paxton2011mesa} and synthetic light curves were obtained by the one-dimensional multi-frequency radiation hydrodynamics code \texttt{STELLA} \cite{blinnikov2000}. In general, the shock breakout \cite{2023PASJ...75..634M} luminosity in the entire early UV light curve is not expected to be as luminous as the plateau optical luminosity, as is observed in \snname. To power this early UV emission and match the model to observations, we embedded the explosion within a spherically symmetric, dense, and confined circumstellar medium \cite{forster2018} that is extended to $10^{15}~\mathrm{cm}$ with an enhanced mass-loss rate of $3\times 10^{-3}~\mathrm{M_\odot~yr^{-1}}$, and a wind velocity of $10~\mathrm{km~s^{-1}}$. 

This simplified CSM prescription does not account for the first two UV light curve points in F110W and F814W, which could be due to the shock breakout emission of a larger radii star, or interaction with an aspherical CSM \cite{Singh2024_aspherical_CSM, Yang2025_aspherical_CSM}. However, due to the uncertainty in absolute magnification and lack of complimentary early-time optical data, we instead focus instead on connecting the rise in the later UV points to the optical data at \spectralphase. Our approach reasonably captures the UV light curve slopes and relative luminosities, and results in a confined CSM mass of 0.2 M$_{\odot}$. This mass is considered a lower limit, as the limited UV data prevent us from knowing whether this UV flux was near maximum during the \hst~window of observations. We obtained synthetic light curves in the observer frame by shifting synthetic spectral energy distributions to $z=5.133\pm0.001$ and convolving them with the filter response functions. Main Text Figure~5
presents an example of a resultant light-curve model for \snname. This iteration is an explosion of a 16 M$_{\odot}$ RSG progenitor having a hydrogen-rich envelope mass of 10.7 M$_{\odot}$ with a radius of 593 R$_{\odot}$. It has the explosion energy of $1.5\times 10^{51}~\mathrm{erg}$ and the $^{56}$Ni mass of 0.1 M$_{\odot}$. The $^{56}$Ni mass is relatively high when compared to a sample of local SNe~II \cite{2019A&A...628A...7A}, but it is close to that estimated for SN~1987A \cite{2014ApJ...792...10S}. Mixing of $^{56}$Ni into the hydrogen-rich envelope affects the light-curve evolution \cite{2019MNRAS.483.1211K}, and we calculate two complimentary models with no mixing and full mixing, to create an envelope that matched \snname' observed color and late-time light curve behavior.

\subsection{Low-$z$ Analog Comparisons} \label{sec:SM_low_z_comparisons}
\noindent \snname~is the farthest spectroscopically classified supernova ever discovered, and is consistent with having a very low metallicity. Thus, we compare it directly to local Universe, low-metallicity (``low-$Z$'') SNe~II which have been suggested to be good analogs to SNe at high-$z$, as well as to a best-matching solar metallicity SN as a control. For the low-$Z$ sample, there have only been three well-studied SNe~II at $Z\lesssim 0.1~Z_{\odot}$: SN~2015bs \cite{anderson_2015bs_2018}, SN~2017ivv \cite{Gutierrez2020_2017ivv}, and SN~2023ufx \cite{tucker_ufx_2024, Ravi2025_ufx}. For our solar-metallicity control, we compare against SN~1992H \cite{1996AJ....111.1286C}, which was our best spectral template match (see Supplementary Text section~\ref{sec:spectral_diffs}). The three low-$Z$ SNe exhibit characteristics that are distinct from their solar-metallicity counterparts. Each SN was associated with a very low mass host galaxy, consistent with their low metallicity, having optical peaks $\sim0.7-1.7$ brighter than the average [$\bar{M}_{B} = -16.80\pm0.37$~AB~mag; \cite{Richardson2014_AbsMag_dist}], as well as faster declining plateau phases ($\sim40-50$ vs $\sim100$~days) than the bulk of the population reported in the literature [see the following non-exhaustive list, \cite{Anderson2014_SNIIP_LCs, Martinez2022_CSP1, Martinez2022_CSP_physical_params, Martinez2022_SNII_diversity}]. In the case of \snname, we find that our model light curve has properties that span this comparison sample, with our inferred peak magnitude being similarly brighter than average ($-17.7$~AB~mag), however, with a long duration plateau more closely matching SN~1992H [as well as other typical, local SNe~II \cite{Martinez2022_CSP1}]. 

Main Text Figure 2 compares \snname~to these four SNe, selected at a similar phases ($\gtrsim$~98~days) to our spectral observation of \snname. The exception is SN~2015bs, which we plot at a phase of $+57$~days from explosion, because the spectrum at $+76$~days did not have high enough signal-to-noise to compare Fe II. Remarkably, we find that \snname~overall compares most favorably to SN~1992H, especially in terms of its \halpha~and Na I doublet (rest-frame $\approx 5892$~\AA) features. The disparity in \halpha~between \snname~and SN~2017ivv may be due to a difference in phase, however, the boxy shape of SN~2023ufx's \halpha~profile may indicate further interaction with an extended medium at \spectralphase [\cite{Dessart2022_SNII_Interaction}, although see the discussion in \cite{tucker_ufx_2024}]. Such features are not apparent in the spectrum of \snname. In the right-hand panel of Main Text Figure 2
, we focus on the Fe II absorption feature of each SN, and find extremely good agreement with all three low-$Z$ SNe in the pEW of Fe~II (including with SN~2015bs despite only comparing the $+57$~day spectrum), whereas the absorption for SN~1992H is much deeper. More detailed comparisons between these SNe and \snname~will require additional data: photometric observations of \snname' decline off of the plateau, as well as nebular spectra, will constrain the amount of $^{56}$Ni synthesized, and therefore the core mass of the progenitor. Combining this with the envelope mass estimate from the plateau duration will enable a more confident inference on the zero-age main sequence mass of the progenitor to \snname, potentially holding implications for the IMF at $z>5$.

\input{Combined_Phot}

\begingroup
\renewcommand{\arraystretch}{1.5}
\begin{table}[t]
    \centering
    \begin{tabular}{lccccccc}
    \hline
    \hline
    Image & $\alpha$ & $\delta$ & $\left|\mu\right|$ & $\tau$\\
     & [deg, J2000] & [deg, J2000] & & [time, observer] \\
    \hline
    101.1 & 292.9550834 & -26.5748399 & $25.46 \pm 5.73$ & $0$ \\
    101.2 & 292.9549276 & -26.5746208 & $29.88 \pm 6.22$ & $0.99 \pm  0.33$ \,days \\ 
    p101.3 & 292.9516835 & -26.5782154 & $12.93 \pm 2.59$ & $-3.68 \pm 0.74$ years\\
    p101.4 & 292.9532987 & -26.5793043 & $13.01 \pm 2.91$ & $-3.49 \pm 0.70$ years\\  
    p101.5 & 292.9683686 & -26.5760254 & $2.15 \pm 0.43$ & $-53.99 \pm 10.80$ years\\
    \hline
    \hline
    \end{tabular}
    
    \caption{\textbf{Summary of the detected and predicted images of SN Eos.} Predicted images are preceded with a `p'. Columns are: (i) Object Id; (ii) \& (iii) R.A. and Dec. in degrees; (iv) Lensing magnification $\mu$ for the MCMC-resultant average; (v) Relative time delay $\tau$ in the observer frame, measured from the arrival of image 101.1, for the MCMC-resultant average. Image 101.2 is estimated to be the last one to arrive. 
    }
    \label{tab:detections}
\end{table}
\endgroup

%% file: Combined_Phot.tex
\begin{table*}
    \centering
    \begin{tabular}{cccccc}
\hline
PID & MJD & Instrument & Filter/Grating & 101.1 (AB mag) &  101.2 (AB mag) \\
\hline
17504 & 60380.1 & \textit{HST}/WFC3-UVIS & \textit{F814W} & 26.40 $\pm$ 0.16 & 26.08 $\pm$ 0.12 \\
17504 & 60380.1 & \textit{HST}/WFC3-IR   & \textit{F110W} & 25.51 $\pm$ 0.05 & 24.81 $\pm$ 0.04 \\
17504 & 60382.1 & \textit{HST}/WFC3-UVIS & \textit{F814W} & 26.61 $\pm$ 0.34 & 26.30 $\pm$ 0.27 \\
17504 & 60382.1 & \textit{HST}/WFC3-IR   & \textit{F110W} & 25.37 $\pm$ 0.06 & 24.83 $\pm$ 0.04 \\
17504 & 60384.1 & \textit{HST}/WFC3-UVIS & \textit{F814W} & 26.23 $\pm$ 0.25 & 25.91 $\pm$ 0.20 \\
17504 & 60384.1 & \textit{HST}/WFC3-IR   & \textit{F110W} & 25.59 $\pm$ 0.12 & 25.04 $\pm$ 0.07 \\
17504 & 60386.2 & \textit{HST}/WFC3-UVIS & \textit{F814W} & 25.63 $\pm$ 0.27 & 26.04 $\pm$ 0.35 \\
17504 & 60386.2 & \textit{HST}/WFC3-IR   & \textit{F110W} & 25.45 $\pm$ 0.15 & 24.74 $\pm$ 0.09 \\
17504 & 60387.0 & \textit{HST}/WFC3-IR   & \textit{F110W} & 25.51 $\pm$ 0.18 & 24.57 $\pm$ 0.08 \\
\hline
6882  & 60919.0 & \textit{JWST}/NIRCam   & \textit{F090W} & $>28.0$          & $>28.0$          \\
6882  & 60919.0 & \textit{JWST}/NIRCam   & \textit{F115W} & $>28.0$          & $>28.0$          \\
6882  & 60919.0 & \textit{JWST}/NIRCam   & \textit{F150W} & $>28.0$          & $>28.0$          \\
6882  & 60919.0 & \textit{JWST}/NIRCam   & \textit{F200W} & $>28.0$          & $>28.0$          \\
6882  & 60919.0 & \textit{JWST}/NIRCam   & \textit{F210M} & $>28.0$          & $>28.0$          \\
6882  & 60919.1 & \textit{JWST}/NIRCam   & \textit{F277W} & 26.80 $\pm$ 0.01 & 26.63 $\pm$ 0.03 \\
6882  & 60919.1 & \textit{JWST}/NIRCam   & \textit{F300M} & 26.27 $\pm$ 0.01 & 26.25 $\pm$ 0.03 \\
6882  & 60919.1 & \textit{JWST}/NIRCam   & \textit{F356W} & 25.94 $\pm$ 0.01 & 25.81 $\pm$ 0.02 \\
6882  & 60919.0 & \textit{JWST}/NIRCam   & \textit{F410M} & 25.27 $\pm$ 0.01 & 25.26 $\pm$ 0.01 \\
6882  & 60919.1 & \textit{JWST}/NIRCam   & \textit{F444W} & 25.44 $\pm$ 0.01 & 25.25 $\pm$ 0.01 \\
\hline
9493  & 60956.6 & \textit{JWST}/NIRCam   & \textit{F070W} & $>27.5$          & $>27.5$          \\
9493  & 60956.6 & \textit{JWST}/NIRCam   & \textit{F277W} & 27.21 $\pm$ 0.01 & 27.11 $\pm$ 0.05 \\
9493  & 60956.6 & \textit{JWST}/NIRCam   & \textit{F356W} & 26.20 $\pm$ 0.01 & 26.07 $\pm$ 0.03 \\
9493  & 60956.6 & \textit{JWST}/NIRCam   & \textit{F444W} & 25.63 $\pm$ 0.01 & 25.51 $\pm$ 0.02 \\
9493  & 60956.6 & \textit{JWST}/NIRSpec  & \textit{PRISM} & --&--    \\
\hline
\end{tabular}
\begin{flushleft}
\caption{
\label{tab:phot} \textbf{All observations of SN Eos.}
Columns are: (i) \textit{HST/JWST} Program ID, (ii) Modified Julian date, (iii) \webb/\hst~instrument, (iv) WFC3/NIRCam filter, (v) \& (vi) photometry from lens image 101.1 and 101.2 for SN Eos in AB mag. Upper limits are 5\,$\sigma$.}
\end{flushleft}
\end{table*}